\documentclass[aps,prb,preprint,superscriptaddress]{revtex4-1}
\usepackage{graphicx}% Include figure files
\usepackage{dcolumn}% Align table columns on decimal point
\usepackage{bm}% bold math
\usepackage[breaklinks]{hyperref}%hyperef package does internal and external linking (labels, urls)
\usepackage[all]{hypcap}%addition to hyperref, it ensures figure is correctly displayed (and not only the caption) when link is clicked
\usepackage{color}
\usepackage{array}
\newcommand{\ff}{\frac{1}{2}}
\newcommand{\nn}{\nonumber}

\begin{document}

%\preprint{}

\title{Non-Lifshitz-Kosevich field- and temperature-dependent amplitude of quantum oscillations in the quasi-two dimensional metal $\theta$-(ET)$_4$ZnBr$_4$(C$_6$H$_4$Cl$_2$) }

%\author{Alain Audouard$^{1}$, David Vignolles$^{1}$, Rustem B. Lyubovskii$^{2}$, Lo\"{\i}c Drigo$^{1}$,  Rimma N. Lyubovskaya$^{2}$, and Enric~Canadell$^{3}$}

%\affiliation{$^1$ Laboratoire National des Champs Magn\'{e}tiques
%Intenses (UPR 3228 CNRS, INSA, UJF, UPS) 143 avenue de Rangueil,
%F-31400 Toulouse,
%\\$^3$Institut de Ci\`{e}ncia de Materials de Barcelona, Consejo
%Superior de Investigationes Cient\'{i}ficas, Campus Universitat
%Aut\`{o}noma de Barcelona, Bellaterra 08193,
%Spain\\$^4$Institute of Problems of Chemical Physics, Russian
%Academy of Sciences, 142432
%Chernogolovka, MD, Russia}%

\author{Alain Audouard}
\affiliation{Laboratoire National des Champs Magn\'{e}tiques
Intenses (UPR 3228 CNRS, INSA, UJF, UPS) 143 avenue de Rangueil,
F-31400 Toulouse, France.}

\author{Jean-Yves~Fortin}
\affiliation{Institut Jean Lamour, D\'epartement de Physique de la
Mati\`ere et des Mat\'eriaux,
CNRS-UMR 7198, Vandoeuvre-les-Nancy, F-54506, France.%\\This line break forced% with \\
}%

\author{David Vignolles}
\affiliation{Laboratoire National des Champs Magn\'{e}tiques
Intenses (UPR 3228 CNRS, INSA, UJF, UPS) 143 avenue de Rangueil,
F-31400 Toulouse, France.}

\author{Rustem~B.~Lyubovskii}
\affiliation{Institute of Problems of Chemical Physics, Russian Academy of Sciences, 142432 Chernogolovka, MD, Russia}%

\author{Lo\"{\i}c Drigo}
\affiliation{Laboratoire National des Champs Magn\'{e}tiques
Intenses (UPR 3228 CNRS, INSA, UJF, UPS) 143 avenue de Rangueil,
F-31400 Toulouse, France.}

\author{Gena V. Shilov }
\affiliation{Institute of Problems of Chemical Physics, Russian Academy of Sciences, 142432 Chernogolovka, MD, Russia}

\author{Fabienne~Duc}
\affiliation{Laboratoire National des Champs Magn\'{e}tiques
Intenses (UPR 3228 CNRS, INSA, UJF, UPS) 143 avenue de Rangueil,
F-31400 Toulouse, France.}

\author{Elena I. Zhilyaeva }
\affiliation{Institute of Problems of Chemical Physics, Russian Academy of Sciences, 142432 Chernogolovka, MD, Russia}

\author{Rimma N. Lyubovskaya}
\affiliation{Institute of Problems of Chemical Physics, Russian Academy of Sciences, 142432 Chernogolovka, MD, Russia}

\author{Enric Canadell}
 \affiliation{Institut de Ci\`{e}ncia de Materials de Barcelona, CSIC, Campus de
la UAB, 08193, Bellaterra, Spain.%\\This line break forced% with \\
}%

\date{\today}

% $\dagger$ author for correspondence: andreev@apollo.karlov.mff.cuni.cz

%

\begin{abstract}

According to band structure calculations, the Fermi surface of the quasi-two dimensional metal $\theta$-(ET)$_4$ZnBr$_4$(C$_6$H$_4$Cl$_2$) illustrates the linear chain of coupled orbits model. Accordingly, de Haas-van Alphen oscillations spectra recorded in pulsed magnetic field of up to 55 T evidence many Fourier components, the frequency of which are linear combinations of the frequencies relevant to the closed $\alpha$ and the magnetic breakdown $\beta$ orbits. The field and temperature dependence of these components' amplitude are quantitatively accounted for by analytic calculations including, beyond the Lifshitz-Kosevich formula, second order terms in damping factors due to the oscillation of the chemical potential as the magnetic field varies. Whereas these second order terms are negligible for the orbits $\alpha$, $\beta$ and  $2\beta-\alpha$, they are solely responsible for the 'forbidden orbit' $\beta-\alpha$ and its harmonic and have a significant influence on Fourier components such as $2\alpha$ and $\beta+\alpha$, yielding strongly non-Lifshitz-Kosevich behaviour in the latter case.


\end{abstract}


\maketitle

\section{Introduction}

\begin{figure}%[h]                                                    % Figure mm_TF_FS
\centering
%\resizebox{\columnwidth}{!}
%\resizebox{12cm}{!}{\includegraphics*{F(q).eps}}
\includegraphics[width=0.7\columnwidth,clip,angle=0]{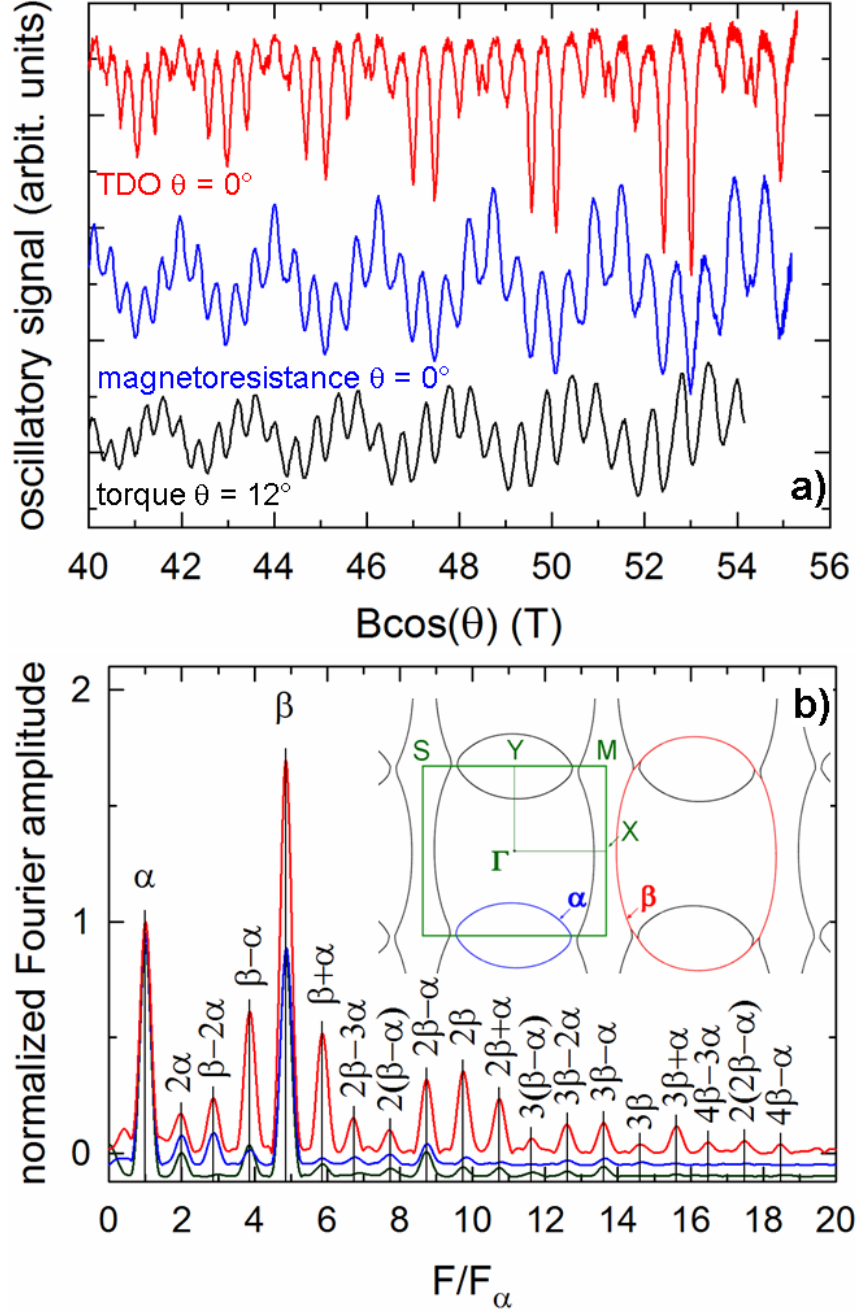}
\caption{\label{Fig:mm_TF_FS} (color on line)  (a) Oscillatory part of the high field range of TDO frequency (crystal \#6), 4-point interlayer magnetoresistance (crystal \#5) and torque (crystal \#3) data at 2 K and (b) corresponding Fourier analysis (Fourier spectra are shifted down from each other by a constant amount for clarity). Thin lines in (b) are marks calculated with $F_{\alpha}$ = 0.93 kT and  $F_{\alpha}/F_{\beta}$ = 0.205. The inset displays the Fermi surface of the conducting layer at 100 K. Green rectangle depicts the first Brillouin zone where $S$  = ($-a^*$/2, $b^*$/2), $Y$  = (0, $b^*$/2), $M$ =  ($a^*$/2, $b^*$/2), $X$  = ($a^*$/2, 0) and $\Gamma$ = (0, 0). The basic orbits $\alpha$ and $\beta$ are marked by the blue and red lines, respectively. }
\end{figure}

Many quasi-two-dimensional (q-2D) metals, in particular charge transfer salts based on the bis-ethylenedithio-tetrathiafulvalene (ET) molecule, illustrate the textbook Fermi surface (FS) proposed by Pippard in the early sixties. This model FS, an example of which is provided in the inset of Fig.~\ref{Fig:mm_TF_FS}, was intended to compute the de Haas-van Alphen (dHvA) oscillation spectrum of the linear chain of orbits coupled by magnetic breakdown (MB) \cite{Pi62,Sh84}. In line with the coupled orbits network model of Falicov-Stachowiak\cite{Fa66,Sh84}, relevant dHvA oscillations spectra involve linear combinations of frequencies linked to the $\alpha$ and MB-induced $\beta$ orbits \cite{Uj97,St99,Ha96b,Ca94,Ka04,Uj08}. However, it is now well established that the field and temperature dependence of many of these Fourier components cannot be accounted for by this model due to oscillation of the chemical potential in magnetic field\cite{Al96,Fo98,Al01,Ch02,Ki02,Gv02,Fo05}.
Analytic tools, given in the Appendix, have been provided in order to quantitatively account for the field- and temperature-dependent amplitudes of the various Fourier components observed \cite{Au12,Au13,Au14}. Briefly, in addition to a first order term corresponding to the Lifshitz-Kosevich (LK) model\cite{Sh84}, second order terms due to oscillation of the chemical potential must be taken into account. Nevertheless, their relative importance strongly depends on the involved parameter values, in particular the Land\'{e} factors. As an example, provided spin damping factors relevant to basic orbits are not too small, i. e. $g^*_{\alpha}m_{\alpha}$ and $g^*_{\beta}m_{\beta}$ (where $g^*_{\alpha(\beta)}$ and $m_{\alpha(\beta)}$ are the effective Land\'{e} factor and effective mass, respectively, of the $\alpha (\beta)$ orbit) are not close to odd integers, these second order terms have a negligible contribution to the Fourier amplitude $A_{\alpha}$ and $A_{\beta}$, respectively. In contrast, the  amplitudes  $A_{p(\beta-\alpha)}$ of the Fourier components with frequencies $p(F_{\beta}-F_{\alpha})$, which are commonly referred to as 'forbidden orbits' since they do not correspond to MB orbits, are only governed by second order terms. For completeness, It should be noticed that, in the case of magnetoresistance oscillations, components such as $\beta-\alpha$ or $\beta-2\alpha$ correspond to quantum interference paths \cite{St71} which are liable to enter the Shubnikov-de Haas (SdH) spectra \cite{Ka96}.

Up to now, these calculations have only been implemented to account for the data of the strongly two-dimensional compound $\theta$-(ET)$_4$CoBr$_4$(C$_6$H$_4$Cl$_2$)\cite{Au12,Au13}, referred in the following to as the Co-compound. For this compound, the field and temperature dependence of the second harmonic amplitude of the $\alpha$ orbit ($A_{2\alpha}$), which significantly differs from the predictions of the LK model, and the 'forbidden orbit' amplitude $A_{\beta-\alpha}$ are quantitatively accounted for by the calculations. Nevertheless, data analysis for other compounds, with different FS parameters, are needed to further check the model. In addition, depending on the value of the involved FS parameters (in particular effective masses and Land\'{e} factors), strongly non-monotonic field and temperature dependence is liable to be observed in few cases \cite{Au12,Au14}. Actually, such a feature has never been reported yet.

The aim of this article is to report on quantum oscillations spectra of $\theta$-(ET)$_4$ZnBr$_4$(C$_6$H$_4$Cl$_2$), referred to as the Zn-compound in the following. This compound belongs to the same family as the Co-compound, namely  $\theta$-(ET)$_4$MBr$_4$(C$_6$H$_4$Cl$_2$), where M is a metal such as Co, Zn, Hg, Cd (for a review, see Ref.~\onlinecite{Ly14}). Strikingly, the crystal structure of these compounds involves one conducting and one insulating ET plane, with different atomic arrangement, insuring a strong two-dimensionality. More extended data than for the previously reported Co-compound, i.e. field and temperature dependence of Fourier amplitude relevant to several frequency combinations, are derived, allowing a more extensive check of the formulas reported in the Appendix. In particular, it is demonstrated that strongly non-monotonic temperature dependence of the Fourier component with frequency corresponding to the MB orbit $\beta$+$\alpha$ is observed.

\section{Experimental}

Crystals were synthesized by electrocrystallization technique as reported in Ref.~\onlinecite{Sh11}. The FS topology was obtained through extended H\"{u}ckel type tight-binding band structure calculations\cite{Wh78}, as reported in Refs.~\onlinecite{Pe93,Au11}. These calculations were based on X-ray diffraction data collected at 100 K and 180 K at the IPCP-Chernogolovka and the LCC-Toulouse, respectively.

Six crystals denoted hereafter as crystal \#1 to \#6, respectively, were studied in pulsed magnetic fields of up to 55 T with a pulse decay duration of 0.32 s. DHvA oscillations were measured through magnetic torque measurements of crystals \#1 to \#4, with approximate dimensions 0.1 $\times$ 0.1 $\times$ 0.04~mm$^3$, stuck on a microcantilever. Variations of the microcantilever piezoresistance were measured at liquid helium temperatures with a Wheatstone bridge with an $ac$ excitation at a frequency of 63 kHz. The angle between the normal to the conducting plane and the magnetic field
direction was $\theta$ = 11$^{\circ}$, 8$^{\circ}$ and 9$^{\circ}$ for crystals \#1, \#2 and \#3, respectively, while $\theta$ was varied from 15$^{\circ}$ to 71$^{\circ}$ thanks to a rotating sample holder for crystal \#4. SdH oscillations were measured through 4-point interlayer magnetoresistance (crystal \#5) and contactless tunnel diode oscillator (TDO)-based method \cite{Dr10,Au12} (crystal \#6).

\section{Results and discussion}

In the next section (\ref{sec:structures}) the oscillatory data are examined at the light of band structure calculations. Section \ref{sec:basic} reports on the field and temperature dependence of the basic Fourier components, linked to the $\alpha$ and $\beta$ orbits, observed in dHvA and SdH spectra while dHvA frequency combinations are considered in Section \ref{sec:combinations}.

\subsection{Band structure calculations and oscillatory spectrum}
\label{sec:structures}

Crystalline and electronic band structures of the Zn-compound are very similar to those of the Co-compound. Briefly, two different cation layers, labeled $A$ and $B$, respectively, in Refs.~\cite{Sh11,Au12,Ly14}, with different atomic arrangements, are observed within the unit cell. According to band structure calculations, layer  $A$ with $\alpha$-type packing is insulating  while layer $B$ with $\theta$-type packing is conducting (for details regarding atomic packing in organic metals, see Ref.~\onlinecite{Sh04}). As reported in Fig.~\ref{Fig:mm_TF_FS}, the FS topology relevant to layer $B$ illustrates the Pippard's model, observed in many organic conductors based on the ET molecule. Namely, it is composed of one q-2D closed tube and two q-1D sheets separated by a gap. In magnetic fields, the closed tube yields the $\alpha$ orbit while, thanks to MB, the $\beta$ orbit with an area equal to that of the first Brillouin zone (FBZ) is observed. The area of the $\alpha$ orbit is 17.0 \% and 18.2\% of the FBZ area at 180 K and 100 K, respectively. It can be remarked that the FS of Fig.~\ref{Fig:mm_TF_FS} differs from that of other $\theta$-phase salts. In these latter salts the anions impose a periodicity along the $b$ direction which is different from that observed in the Zn- and Co-compounds, yielding different FS topology\cite{Ko86}.

Fig.~\ref{Fig:mm_TF_FS}(a) displays oscillatory parts of the magnetic torque, 4-point longitudinal magnetoresistance and TDO data at 2 K. It can be remarked first that, while TDO and 4-point magnetoresistance data are in phase, magnetic torque data are phase-shifted by $\pi$/2. This feature indicates  that, while magnetic torque yields dHvA oscillations, both TDO and 4-point magnetoresistance yield SdH oscillations, in agreement with previous statements \cite{Co00,Oh04}. Corresponding Fourier analysis are displayed in Fig.~\ref{Fig:mm_TF_FS}(b). The two main frequencies, $F_{\alpha}$ = 0.930(2) kT and $F_{\beta}$ = 4.534(7) kT,  correspond to the $\alpha$ and $\beta$ orbits, respectively, hence the $\alpha$ orbit area amounts to 20.5 \% of the FBZ area. This value is in good agreement with the above reported band structure calculations which are based on X-ray diffraction data measured at higher temperature, owing to the increase of the closed tube area relatively to that of the FBZ, as the temperature decreases from 180 K to 100 K. Strikingly, an unprecedented number of frequencies is observed, in particular in the case of the TDO data, accounting for the strong non-sinusoidal oscillatory part of these data. These frequencies, labeled $F_{\eta}$ in the following, are linear combinations of $F_{\alpha}$ and $F_{\beta}$. Frequency as high as 17.2 kT, corresponding to $\eta = 4\beta-\alpha$, is observed in the TDO spectrum of Fig.~\ref{Fig:mm_TF_FS}.

\subsection{Basic Fourier components amplitude}
\label{sec:basic}

\begin{figure}%[h]                                                    % Figure 2 mass_plot_alpha_beta
\centering
%\resizebox{\columnwidth}{!}
%\resizebox{12cm}{!}{\includegraphics*{F(q).eps}}
\includegraphics[width=0.6\columnwidth,clip,angle=0]{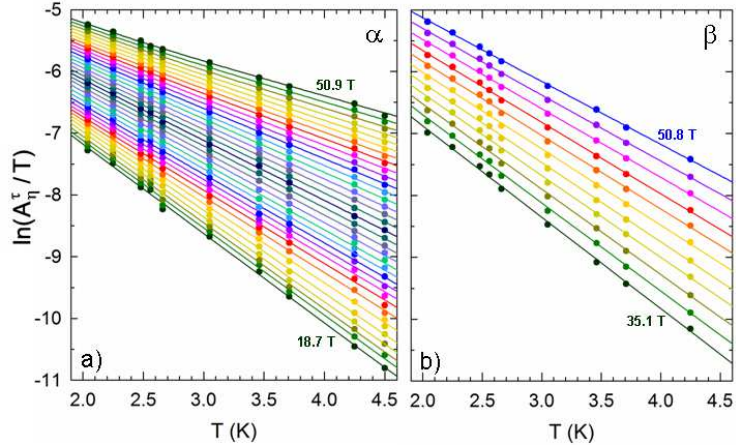}
\caption{\label{Fig:mass_plot_alpha_beta} (color on line) Temperature dependence (mass plots) of the magnetic torque amplitudes (a) $A^{\tau}_{\alpha}$ and (b) $A^{\tau}_{\beta}$. Solid lines  in (a) and (b) are best fits of Eqs.~\ref{Eq:alpha} and~\ref{Eq:beta}, respectively, to the data. They are obtained with $m_{\alpha}$= 1.85, $m_{\beta}$= 3.4, $B_0$ = 26 T and $T_D$ = 0.8 K. The considered magnetic field values are evenly spaced in 1/B in the explored field range, the boundary of which are indicated in the figures. }
\end{figure}

\begin{figure}%[h]                                                    % Figure spin_0
\centering
%\resizebox{\columnwidth}{!}
%\resizebox{12cm}{!}{\includegraphics*{F(q).eps}}
\includegraphics[width=0.6\columnwidth,clip,angle=0]{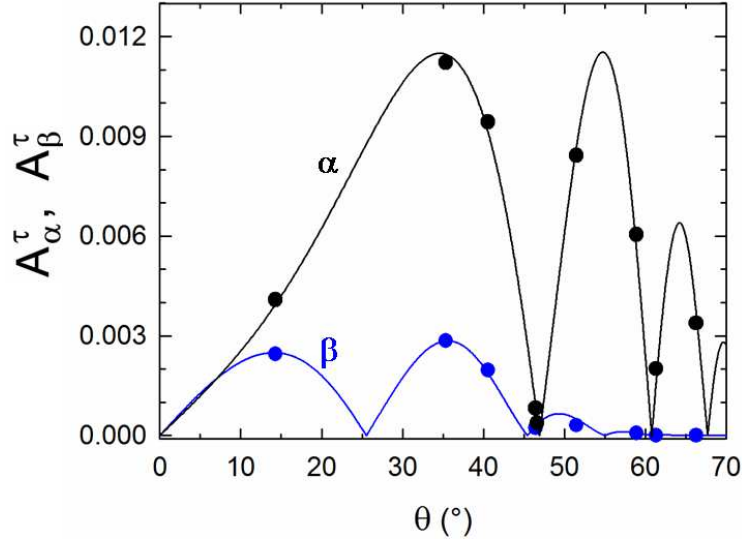}
\caption{\label{Fig:spin_0} (color on line) Angle dependence of the magnetic torque amplitudes $A^{\tau}_{\alpha}$ and $A^{\tau}_{\beta}$. Solid lines are best fits of Eqs.~\ref{Eq:alpha} and Eq.~\ref{Eq:beta} to the data for  $A^{\tau}_{\alpha}$ and $A^{\tau}_{\beta}$, respectively. They are obtained with the same effective mass and MB field as in Fig.~\ref{Fig:mass_plot_alpha_beta} and $g^{*}_{\alpha}$ =  $g^{*}_{\beta}$ = 1.85. }
\end{figure}

\begin{figure}%[h]                                                    % Figure m*(B)  mc_alpha_beta
\centering
%\resizebox{\columnwidth}{!}
%\resizebox{12cm}{!}{\includegraphics*{F(q).eps}}
\includegraphics[width=0.6\columnwidth,clip,angle=0]{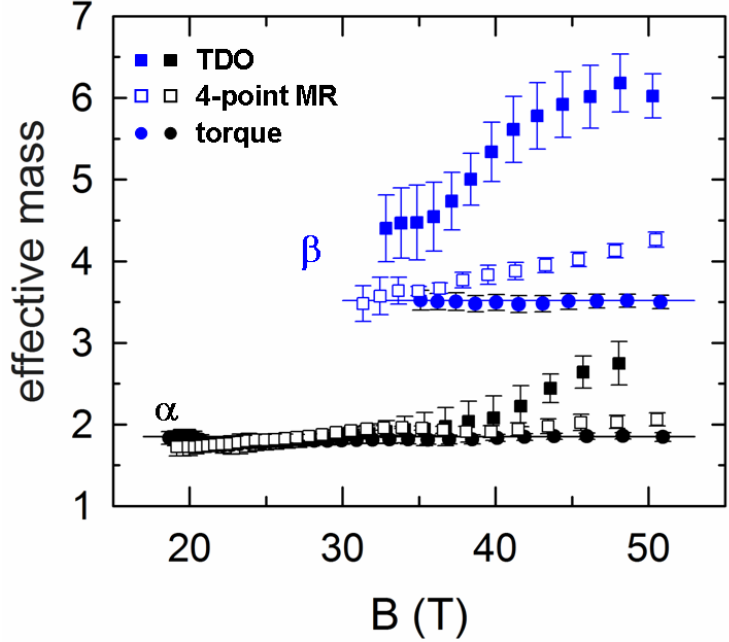}
\caption{\label{Fig:mc_alpha_beta} (color on line) Field dependence of the effective mass value of $\alpha$ and  $\beta$ orbits derived from magnetic torque, 4-point magnetoresistance and TDO data. Horizontal lines mark the effective mass values deduced from magnetic torque data, yielding dHvA oscillations, $m_{\alpha}$ = 1.85 and $m_{\beta}$ = 3.4. }
\end{figure}

Let us consider first magnetic torque data, the oscillation spectra of which involve Fourier components with frequencies $F_{\eta}$ and amplitudes $A^{\tau}_{\eta}$. Since we are dealing with a 2D metal, these amplitudes are related to dHvA oscillations amplitudes $A_{\eta}$ as $A_{\eta}$ $\propto$ $A^{\tau}_{\eta}$/B$\tan(\theta$) where $\theta$ is the angle between the magnetic field direction and the normal to the conducting plane. According to the LK formula, $\ln(A^{\tau}_{\eta}/T)$ is predicted to vary linearly with the temperature at a given magnetic field value (mass plot) at high $T/B$ ratio. Since crystals  \#1,  \#2 and  \#3 yield same results within the error bars reported below, we focus in the following on the data relevant to crystal  \#3, with the lowest Dingle temperature. Data for $\eta$ = $\alpha$ and $\beta$ are reported in Fig.~\ref{Fig:mass_plot_alpha_beta}. They can be analyzed through Eqs.~\ref{Eq:alpha} and~\ref{Eq:beta}, respectively. However, no fewer than seven parameters enter these equations: effective masses $m_{\alpha(\beta)}$, Dingle temperatures $T_{D\alpha(\beta)}$, effective Land\'{e} factors $g^*_{\alpha(\beta)}$ and MB field $B_0$. Nevertheless, as observed in the case of the Co-compound \cite{Au12} and discussed in Ref.~\onlinecite{Au13}, the second order terms of Eqs.~\ref{Eq:alpha} and~\ref{Eq:beta} are negligibly small compared to their leading terms, provided the spin damping factors $R^s_{\alpha(\beta)}$ are far enough from spin-zeroes. As a result, the LK model applies and the spin damping factors act as field- and temperature-independent prefactors. However, in addition to the effective masses, the MB field $B_0$ and the Dingle temperatures $T_{D\alpha(\beta)}$ govern the field dependence of $A_{\alpha(\beta)}$. As a result, each of the two equations~\ref{Eq:alpha} and~\ref{Eq:beta} still involve 3 parameters yielding large uncertainties. For this reason, it is assumed in the following that $T_{D\alpha}$ = $T_{D\beta}$. Within this assumption, data yield  $m_{\alpha}$= 1.85(10), $m_{\beta}$= 3.40(15) and $B_0$ = 26(3) T for all the three studied crystals. The Dingle temperature, which is the only crystal-dependent parameter is $T_{D1}$ = 0.9(1) K,  $T_{D2}$ = 1.1(1) K and $T_{D3}$ = 0.8(1) K for crystal \#1,  \#2 and  \#3, respectively. Effective Land\'{e} factors $g_{\alpha(\beta)}^*$, which are the remaining parameters to be determined, are obtained through the angle dependence of $A_{\alpha(\beta)}$. Solid lines in Fig.~\ref{Fig:spin_0} are the best fits of Eqs.~\ref{Eq:alpha} and~\ref{Eq:beta} to the data relevant to crystal \#4, yielding $g^{*}_{\alpha}$ =  $g^{*}_{\beta}$ = 1.85(10).

In short, the effective masses and Dingle temperatures of the Zn-compound are close to the data obtained for the Co-compound whereas the MB field of the latter is higher \cite{Au12}. Owing to the effective Land\'e factors values, which were only estimated in Ref.~\onlinecite{Au12}, the second order terms of Eqs.~\ref{Eq:alpha} and~\ref{Eq:beta} are negligible in the field and temperature range explored, indicating that the LK model, i.e. the first order term of Eqs.~\ref{Eq:alpha} and~\ref{Eq:beta} satisfactorily accounts for the basic orbits $\alpha$ and $\beta$, respectively.

A hallmark of the validity of the LK formula is the field-independency of the effective mass derived through this formula from the temperature dependence of the amplitude, as it can be observed in Fig.~\ref{Fig:mc_alpha_beta} for the dHvA data. In contrast, an apparent strong increase of the effective mass is observed in the case of TDO and, to a less extent, of the 4-point interlayer magnetoresistance data. This behaviour can be ascribed to the failure of the LK formula for SdH oscillations relevant to basic orbits of q-2D metals \cite{La95,Sa96,Ha96}. This feature, which is beyond the scope of the present study focused on dHvA spectra, requires specific calculations of the conductivity  \cite{Gr03,Th08,Be09,En09} taking into account the multiband nature of the FS. As for the smaller discrepancy observed for 4-point magnetoresistance compared to contactless TDO measurements, it must be considered that interlayer resistance ($R_{zz}$) and in-plane resistance ($R_{xx}$) which are governed by different matrix elements \cite{En09}, are measured in the former and latter case, respectively. Besides, the electrical contacts on the crystal, in the case of 4-point magnetoresistance, connect the quasi-particles to a non-quantized reservoir, liable to induce a damping of the chemical potential oscillation\cite{Ha96}.

\subsection{Frequency combinations}
\label{sec:combinations}

\begin{figure}%[h]                                                    % Figure mass_plot_combinaisons
\centering
%\resizebox{\columnwidth}{!}
%\resizebox{12cm}{!}{\includegraphics*{F(q).eps}}
\includegraphics[width=0.6\columnwidth,clip,angle=0]{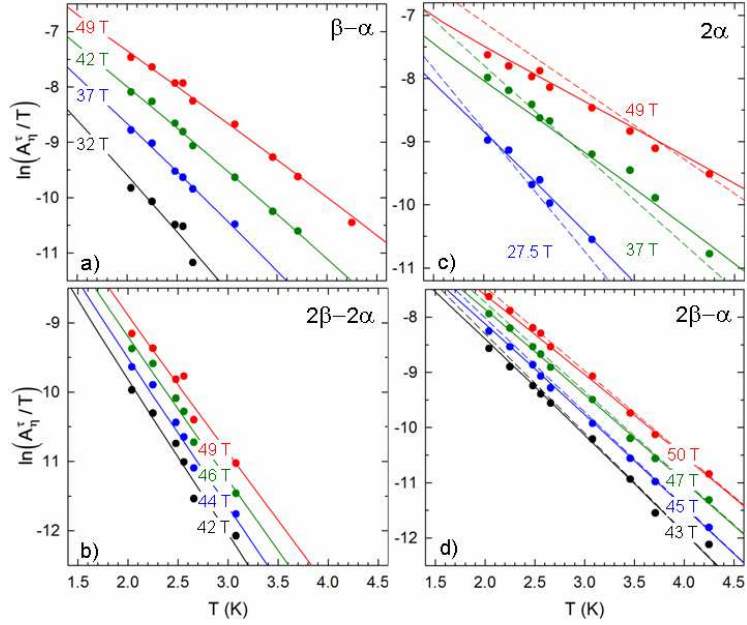}
\caption{\label{Fig:mass_plot_combinaisons} (color on line)  Mass plot of magnetic torque amplitudes for different values of the magnetic field of (a) $\beta-\alpha$, (b) $2(\beta-\alpha)$, (c) $2\alpha$ and (d) $2\beta-\alpha$. Solid lines in (a), (b), (c) and (d) are calculated with Eqs.~\ref{Eq:beta-alpha},~\ref{Eq:2beta-2alpha},~\ref{Eq:2alpha} and ~\ref{Eq:2beta-alpha}, respectively. They are obtained with the same set of parameters as in Figs.~\ref{Fig:mass_plot_alpha_beta}, \ref{Fig:spin_0}. Dashed lines in (c) and (d) are obtained with the Lifshitz-Kosevich formula. }
\end{figure}

\begin{figure}%[h]                                                    % Figure TF_mass_plot_bplusa
\centering
%\resizebox{\columnwidth}{!}
%\resizebox{12cm}{!}{\includegraphics*{F(q).eps}}
\includegraphics[width=0.6\columnwidth,clip,angle=0]{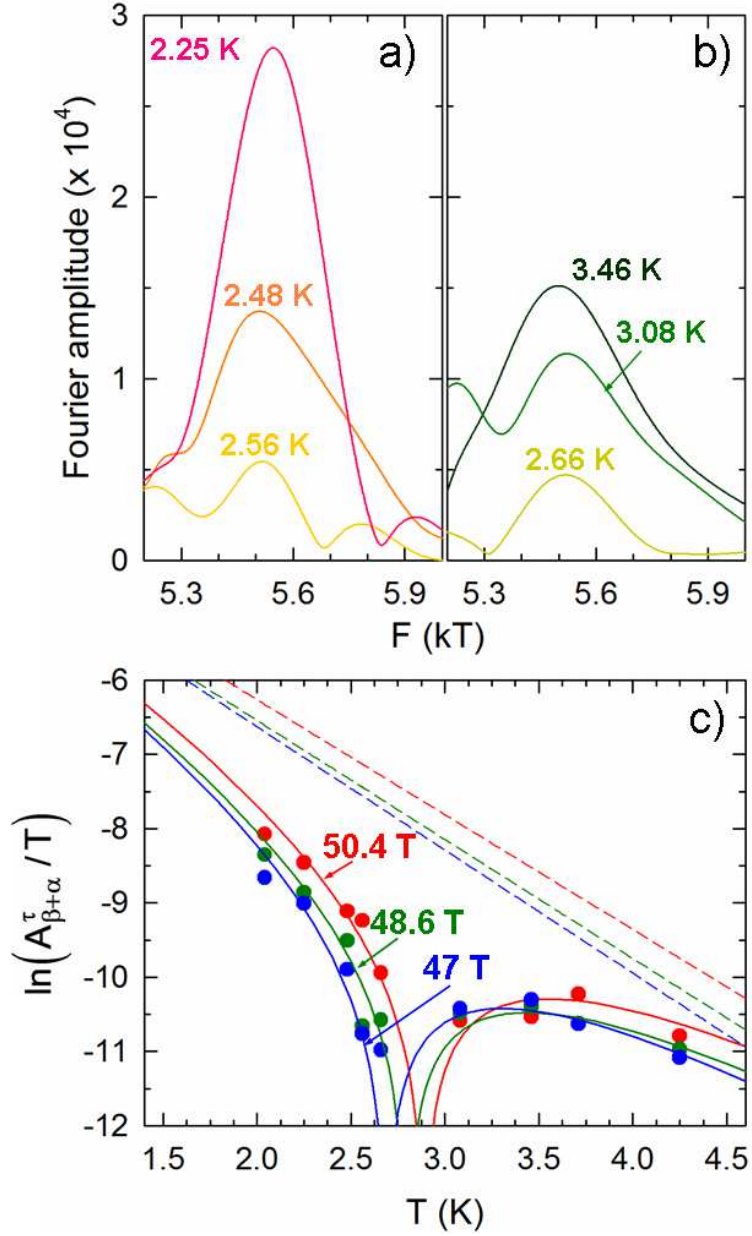}
\caption{\label{Fig:TF_mass_plot_bplusa} (color on line)  Fourier analysis in the frequency range around $F_{\beta+\alpha}$, at the mean magnetic field value $\overline{B}$ = 47 T. In the temperature range (a) 2.25-2.56 K and (b) 2.66-3.48 K, the Fourier amplitude decreases and increases, respectively, as the temperature increases. (c) Mass plot for different values of the magnetic field. Solid lines are calculated with Eq.~\ref{Eq:beta+alpha}. They are obtained with the same set of parameters as in Figs.~\ref{Fig:mass_plot_alpha_beta}, \ref{Fig:spin_0}, \ref{Fig:mass_plot_combinaisons}. Dotted lines are the contributions of the first order term, corresponding to the Lifshitz-Kosevich formula.}
\end{figure}

Since the parameters entering Eqs.~\ref{Eq:alpha} to \ref{Eq:2beta-alpha} are determined from the analysis of the data relevant to the $\alpha$ and $\beta$ orbits, the field and temperature dependence of all the other Fourier components amplitude should be accounted for by these parameters. As examples, the behaviour of few of these amplitudes is considered in Fig.~\ref{Fig:mass_plot_combinaisons}: a very good agreement is indeed observed.

Let us examine these data in more details. First, since the  $\beta-\alpha$ amplitude is dominated by the product $R_{\alpha,1}R_{\beta,1}$ (see Eq.~\ref{Eq:beta-alpha}) its temperature dependence displays a slightly negative curvature. Analyzed through the LK formula, the data would yield an apparent effective mass $m^{app}_{\beta-\alpha}$ close to that of $\beta+\alpha$, actually about 0.8$(m_{\beta}+m_{\alpha})$. This result is in agreement with both experimental data relevant to $\kappa$-ET$_2$Cu(NCS)$_2$ for which $m^{app}_{\beta-\alpha} \simeq$ 0.85$(m_{\beta}+m_{\alpha})$ \cite{Uj97}, and in line with numerical simulations \cite{Fo05}.

As already observed \cite{Uj97,Uj08,Wo08}, $2\alpha$ is not accounted for by the LK formula as well. Indeed, according to Eq.~\ref{Eq:2alpha}, the second order term  which is of the same order of magnitude as the LK damping factor $R_{\alpha,2}$ is dominated by $R_{\alpha,1}^2$ accounting for a non-LK behaviour.

Oppositely, the amplitude of the $2\beta-\alpha$ component is very close to the prediction of the LK model since the second order term, dominated by the product $R_{\alpha,1}R_{\beta,2}$ is very small compared to the LK term which is proportional to $R_{2\beta-\alpha,1}$.

Finally, the Fourier component $\beta+\alpha$ is considered in Fig.~\ref{Fig:TF_mass_plot_bplusa}. While below about 2.6 K, the amplitude decreases as the temperature decreases, it increases in the range 2.6 K $\sim$ 3.5 K, in strong discrepancy with the LK formula. This behaviour is quantitatively well accounted for by Eq.~\ref{Eq:beta+alpha} which evidences a dip in the temperature dependence of the amplitude. Indeed, the second order term of Eq.~\ref{Eq:beta+alpha} is dominated by the product  $R_{\alpha,1}R_{\beta,1}$ which contributes to the amplitude with an opposite sign to the first order Lifshitz-Kosevich term proportional to $R_{\alpha+\beta,1}$. These two factors cancel each other at a given field and temperature value (e.g. 2.9 K at 50 T in the present case), depending on the spin damping factors value, hence on the respective values of the products $g^*_{\alpha}m_{\alpha}$ and $g^*_{\beta}m_{\beta}$. It can be remarked that this feature is not observed in the Co-compound \cite{Au12}. Indeed, owing to slightly different effective masses and effective Land\'e factors, the dip in the $\beta+\alpha$ amplitude would be observed around 9 K, i.e. beyond the temperature range in which oscillations can be observed \cite{Au14}.

\section{Summary and conclusion}

Band structure calculations relevant to the quasi-two dimensional metal $\theta$-(ET)$_4$ZnBr$_4$(C$_6$H$_4$Cl$_2$) indicate that this compound illustrates the linear chain of coupled orbits model proposed by Pippard \cite{Pi62,Sh84} (see Fig.~\ref{Fig:mm_TF_FS}) as it is the case for many organic conductors based on the ET molecule. In line with this statement, quantum oscillations spectra evidence many Fourier components, the frequency of which are linear combinations of the frequencies relevant to the closed $\alpha$ and the magnetic breakdown $\beta$ orbits. The field and temperature dependence of the de Haas-van Alphen amplitude of these components is quantitatively accounted for by the analytic calculations reported in the Appendix. Beyond the Lifshitz-Kosevich formula, they include second order terms arising from the chemical potential oscillations. These second order terms have negligible contributions to the amplitude of the basic $\alpha$ and $\beta$ components allowing the determination of the various physical parameters entering the data (effective masses, magnetic breakdown fields, etc.). They have also a minor contribution on the magnetic breakdown orbit $2\beta-\alpha$. Oppositely, they have significant contribution to 2$\alpha$ and $\beta+\alpha$. Although this latter component physically corresponds to a magnetic breakdown orbit, its temperature dependence evidences a strong dip due to the cancelation of the first and second order terms. Finally, the 'forbidden frequency' $\beta-\alpha$ and its harmonic $2\beta-2\alpha$, which are due to the oscillation of the chemical potential, are accordingly accounted for by second order terms, only.

\acknowledgements
Part of the X-ray diffraction experiments have been performed at the  Laboratoire de Chimie de coordination of Toulouse with the help of Laure Vendier. The support of the European Magnetic Field Laboratory (EMFL) is acknowledged. Work in Bellaterra was supported by MINECO through grant FIS2012-37549-C05-05 and Generalitat de Catalunya (2014SGR301). Work in Chernogolovka  was supported by grant of Presidium RAS 1.1.1.9.

\appendix*
\section{Analytical expressions of Fourier amplitudes}
\label{analytical}

In this appendix, are recalled the analytical equations for de Haas-van Alphen amplitudes $A_{p\eta}$ with frequencies $pF_{\eta}$ given in Refs.~\cite{Au12,Au13,Au14}. They are relevant to two-dimensional FS illustrating the Pippard model in which the component $\eta$ is a linear combination of the $\alpha$ and $\beta$ orbits and $p$ is the harmonic order (see insert of Fig.~\ref{Fig:mm_TF_FS}).

\begin{eqnarray}
\label{Eq:alpha}
A_{\alpha}&=&-\frac{F_{\alpha}}{\pi m_{\alpha}}R_{\alpha,1}-
\frac{F_{\alpha}}{\pi m_{\beta}}
\left [
\ff R_{\alpha,1}R_{\alpha,2}
+\frac{1}{6}R_{\alpha,2}R_{\alpha,3}+2R_{\beta,1}R_{\alpha+\beta,1}
+\ff R_{\beta,2}R_{2\beta-\alpha,1}
\right ]
\\ %\nn
\label{Eq:2alpha}
A_{2\alpha}&=&-\frac{F_{\alpha}}{2\pi m_{\alpha}}R_{\alpha,2}+
\frac{F_{\alpha}}{\pi m_{\beta}}\left [
R_{\alpha,1}^2-\frac{2}{3}R_{\alpha,1}R_{\alpha,3}-R_{\alpha,2}
R_{\alpha+\beta,2} \right ]
\\ %\nn
\label{Eq:beta}
A_{\beta}&=&-\frac{F_{\beta}}{\pi m_{\beta}}R_{\beta,1}-
\frac{F_{\beta}}{\pi m_{\beta}}
\left [
\ff R_{\beta,1}R_{\beta,2}
+\frac{1}{6}R_{\beta,2}R_{\beta,3}+2R_{\alpha,1}R_{\alpha+\beta,1}
+2R_{\beta,1}R_{2\beta,1}
\right ]
\\ \nn
\label{Eq:2beta}
A_{2\beta}&=&-\frac{F_{\beta}}{2\pi m_{\beta}}\left
[R_{\beta,2}+2R_{2\beta,1}\right ]+
\frac{F_{\beta}}{\pi m_{\beta}}
\left [
R_{\beta,1}^2-\frac{2}{3}R_{\beta,1}R_{\beta,3}-\frac{1}{4}R_{\beta,2}R_{\beta,4
}-R_{\alpha,2}R_{\alpha+\beta,2}\right .
\\ %\nn
&+&\left .2R_{\alpha,1}R_{2\beta-\alpha,1}-R_{\beta,2}R_{
2\beta,2}-R_{\beta,4}R_{2\beta,1}
\right ]
\\ %\nn
\label{Eq:beta-alpha}
A_{\beta-\alpha}&=&-\frac{F_{\beta-\alpha}}{\pi m_{\beta}}\left [
R_{\alpha,1}R_{\beta,1}+R_{\alpha,2}R_{\alpha+\beta,1}+R_{\beta,2}R_{
\alpha+\beta,1}+R_{\beta,1}R_{2\beta-\alpha,1}\right
]
\\ \nn
\label{Eq:2beta-2alpha}
A_{2(\beta-\alpha)}&=&-\frac{F_{2\beta-2\alpha}}{\pi m_{\beta}}
\left [2R_{\alpha,2}R_{2\beta,1}+2R_{2\beta-\alpha,2}R_{2\beta,1}
+2R_{\alpha,1}R_{2\beta-\alpha,1}+\frac{1}{2}R_{\alpha,4}R_{\alpha+\beta,2}
\right .
\\
&+&\left .
\frac{1}{2}R_{\alpha,2}R_{\beta,2}+\frac{1}{2}R_{\beta,2}R_{2\beta-\alpha,2}
\right ]
\\ %\nn
\label{Eq:beta+alpha}
A_{\beta+\alpha}&=&-\frac{2F_{\beta+\alpha}}{\pi m_{\beta+\alpha}}
R_{\beta+\alpha,1}+\frac{F_{\beta+\alpha}}{\pi
m_{\beta}}\left [R_{\alpha,1}R_{\beta,1}-2R_{\alpha+\beta,2}R_{\alpha+\beta,1}
-\frac { 1}{3}R_ { \beta,3}R_{2\beta-\alpha,1} \right
]
\\
\label{Eq:2beta-alpha}
A_{2\beta-\alpha}&=&-\frac{F_{2\beta-\alpha}}{\pi m_{2\beta-\alpha}}
R_{2\beta-\alpha,1}-
\frac{F_{2\beta-\alpha}}{\pi m_{\beta}}
\left [\ff R_{\alpha,1}R_{\beta,2}+
\frac{1}{3}
R_{\alpha,3}R_{\alpha+\beta,2}\right ]
\end{eqnarray}

Damping factors are given by the LK and coupled orbits network models \cite{Sh84} as $R_{\eta,p}(B,T)$ = $R^T_{\eta,p}(B,T) R^{D}_{\eta,p}(B) R^{MB}_{\eta,p}(B) R^{s}_{\eta,p}$ \cite{Fa66,Sh84} where the temperature, Dingle, MB and spin damping factors are expressed as  $R^{T}_{\eta,p} = pX_{\eta} \sinh^{-1}(pX_{\eta})$, $R^{D}_{\eta,p} = \exp(-pu_0m_{\eta}T_D(B\cos\theta)^{-1})$, $R^{MB}_{\eta,p} = (ip_0)^{n^t_{\eta}}(q_0)^{n^r_{\eta}}$, $R^{s}_{\eta,p} = \cos(\pi g^*_{\eta}m_{\eta}/2\cos\theta)$, respectively. The field-and temperature-dependent variable ($X_{\eta}$) and the constant ($u_0$) are expressed as $X_{\eta}$ = $u_0 m_{\eta} T/(B\cos\theta)$ and $u_0$ = 2$\pi^2 k_B m_e(e\hbar)^{-1}$ = 14.694 T/K. The tunneling ($p_0$) and reflection ($q_0$) probabilities are given by $p_0$ = $e^{-B_0/2B\cos\theta}$ and $p_0^2$ + $q_0^2$ = 1. $T_D$ is the Dingle temperature
defined by $T_{D}$ = $\hbar(2\pi k_B\tau)^{-1}$, where $\tau^{-1}$ is the scattering rate, $B_0$ is the MB field, $m_{\eta}$ and $g^*_{\eta}$ are the effective masses and effective Land\'{e} factor, respectively. It can be noticed that the terms of first order in damping factors correspond to the LK model.

%\bibliography{pseudo_kappa}

 \end{document}